# Wireless whispering-gallery-mode sensor for thermal sensing and aerial mapping

Xiangyi Xu[1], Weijian Chen[1], Guangming Zhao[1], Yihang Li[1], Chenyang Lu[2] and Lan Yang[1*]

[1]Department of Electrical and Systems Engineering, Washington University, St. Louis, MO 63130, USA

[2]Department of Computer Science and Engineering, Washington University, St. Louis, MO 63130, USA

*yang@seas.wustl.edu

**Internet of Things (IoT) [1, 2] employs a large number of spatially distributed wireless sensors to monitor physical environments, e.g., temperature, humidity, and air pressure, have found wide applications including environmental monitoring [3], health care monitoring [4], smart cities [5] and precision agriculture [6]. A wireless sensor can collect, analyze, and transmit measurements of its environment [1, 2]. To date, wireless sensors used in IoT are predominately based on electronic devices that may suffer from electromagnetic interference in many circumstances. Immune to the electromagnetic interference, optical sensors provide a significant advantage in harsh environments [7]. Furthermore, by introducing optical resonance to enhanced light-matter interactions, optical sensors based on resonators exhibit small footprints, extreme sensitivity and versatile functionalities [8, 9], which can significantly enhance the capability and flexibility of wireless sensors. Here we provide the first demonstration of a wireless photonic sensor node based on whispering-gallery-mode (WGM) optical resonators. The sensor node is controlled via a customized iOS app. Its performance was studied in two practical scenarios: (1) real-time measurement of air temperature over 12 hours and (2) aerial mapping of temperature distribution by a sensor node mounted on an unmanned drone. Our work demonstrates the capability of WGM optical sensors in practical applications and may pave the way for large-scale deployments of WGM sensors in IoT.**

High-quality WGM optical resonators that confine light in a small volume via total internal reflection can significantly enhance light-matter interactions, which has benefited a number of applications including microlasers [10, 11], opto-mechanics [12, 13], and non-Hermitian optics [14,

15, 16]. When subject to environmental changes, WGM resonators will experience changes in their spectral properties, e.g., frequency shift/split, linewidth broadening. Based on such mechanisms, WGM resonators have been demonstrated for various sensing applications ranging from thermal sensor [17], humidity sensor [18], magnetometer [19] to nanoparticle/biomolecule detection [20] and atomic ion detection [21]. The successful laboratory demonstrations have encouraged and advanced the practical applications of WGM sensors. However, a system must address two open challenges to fully release the power of the resonator technology for practical applications: (1) the stability of the photonic resonator and its coupler such as fiber-taper waveguide, and (2) the miniaturization of bulky laboratory measurement systems. Several pioneering works along this direction have been demonstrated recently, e.g., WGM optical gyroscope [22] and phone-sized WGM sensing system [23], which integrate single WGM sensor together with its coupler, laser, photodetector, and associated control components into a portable device. The potential of WGM sensors can be further exploited through the integration with wireless interfaces as parts of an IoT system.

In this work, we demonstrated a wireless WGM sensor node that can be integrated into IoT. We also developed a customized iOS app for remote system control, collection and analysis of sensing signals. Through this app, the spectral properties of WGM sensors can be monitored in real time. As the key elements in WSNs, sensor nodes should have the capability to collect sensing signals, perform signal analysis, and communicate with other sensor nodes or the gateway sensor node. The architecture of our wireless WGM sensor node is shown in **Fig. 1(a)**, consisting of sensing module, microcontroller, Wi-Fi unit, and its power supply [23].

In the sensing module, a tunable single-mode distributed Bragg reflector (DBR) laser is used to probe a packaged WGM sensor [24, 25]; the output from the sensor is received by a photodiode detector. The operation of the DBR laser is controlled via its current driver and thermo-electric cooler (TEC). A transimpedance amplifier (TIA) circuit is associated with the photodiode detector and converts the photodiode current into voltage output with proper gain. An ARM Cortex-M3 processor serves as the microcontroller with main functions of laser control (e.g., voltage, current, temperature) and acquisition of transmission spectra of the WGM sensor. The communication between the sensor node and smartphones is via the Wi-Fi unit. It helps to transmit the sensing signals and receive the commands from the customized app. By connecting the Wi-Fi unit to the internet, a worldwide, real-time control of this system can be realized. We also include a system monitor circuit in the mainboard that monitors the key parameters such as voltage of the power supply and the microcontroller, laser diode voltage/current/temperature. A full-sized view of the mainboard is provided in **Fig. 1(c)**.

The sensor used in this system is a packaged WGM microtoroid resonator. It is fabricated by using UV curable low-index polymer to package a microtoroid together with its fiber-taper waveguide [24, 25]. Such a packaged WGM sensor has high quality factor and long-term stability. Light from the DBR laser with central wavelength of 976 nm and linewidth of 10 MHz is sent into the packaged WGM sensor and then received by the photodetector. The frequency of laser light can be tuned by adjusting the laser current and TEC temperature with tuning coefficients of 0.002 nm/mA and 0.07 nm/°C, respectively. By applying a saw-tooth wave with amplitude of 40 mA to the laser diode around a fixed central current, the frequency of laser light is linearly scanned to obtain the transmission spectrum of the WGM sensor.

The interface of the customized iOS app is shown in **Fig. 1(b)**. The app can monitor the key parameters of the system in real time, and remotely control the mainboard, such as setting the laser diode current and temperature, tuning the laser frequency. It can also receive the transmission spectra of the WGM sensor through the Wi-Fi unit, with a waveform update rate of 50 frames per second. In addition, through the integrated mathematical algorithm, the app can perform real-time analysis such as measurement of resonance frequency, linewidth, and quality factor. Detailed information of this iOS app as well as a step-by-step guide is provided in supplementary material.

We first characterized the spectral properties of the packaged WGM sensor by using the app. Its transmission spectrum is shown in **Fig. 2(a)**, with a frequency spanning of 450 GHz. Multiple resonance modes with different resonance frequencies, quality factors and polarizations were observed. The resonance mode with higher quality factor will help to resolve the smaller frequency shift, subsequently improving the sensing performance. The transmission spectrum of a high-quality mode together with a Lorentzian fit is given in **Fig. 2(b)**, with quality factor about

$4.2\times10^5$. To verify the stability of the whole system, the time trace of the linewidth of a resonance mode was recorded for fifteen minutes. An average linewidth of 3.15 GHz together with a standard deviation of 0.03 GHz was observed, corresponding to standard deviation about 1.0%.

With this wireless sensing system, we performed a 12-hour real-time measurement of air temperature on Jun. 18$^{th}$, 2017 in St. Louis, MO, USA. The whole system was mounted on the outside wall of a building. The packaged WGM sensor was in full contact with the surrounding air and was refrained from direct irradiance from sunlight. The optical fibers connecting the packaged sensor with the mainboard were carefully mounted to avoid polarization variations during the measurement. The variation of the resonance frequency induced by the air temperature change was monitored via the customized app. For comparison, we also mounted a commercial thermometer together with the packaged sensor. Through a 12-h measurement, we acquired a plot of frequency shift of the selected resonance mode. As shown in **Fig. 3**, the resonance frequency shift of the packaged WGM sensor matches well with the results from the commercial thermometer.

Introducing mobility to wireless sensor nodes can improve the capability and flexibility of WSNs and helps to meet the needs of certain scenarios with complex dynamic changes [26]. Here we used an unmanned drone to carry the whole system to measure the temperature distribution in a selected area of a city park in St. Louis on June 1, 2017 (See **Fig. 4(a)**). A commercial thermometer with Bluetooth connection was mounted together with the packaged WGM sensor for comparison. The flight path of the unmanned drone was shown in **Fig. 4(b)**, with the starting and ending locations marked. When the drone flied from one measurement location to another, the resonance frequency of WGM sensor shifted due to the temperature variations. The variation of resonance frequency is shown in **Fig. 4(b)**, where the temperature gradient can be clearly seen. The measurements match well with the results from commercial thermometer (**Fig. 4(c)**). A video demo is provided in the supplementary material where the drone carrying the whole system flied from one location with higher temperature to another one with lower temperature. A resonance frequency shift can be clearly observed from the customized app.

In summary, we have demonstrated a wireless WGM sensor node and exploited it for the applications of thermal sensing and aerial mapping. A customized iOS app enables us to monitor and control the system parameters, acquire and analyze the sensing signals in real time. Two application scenarios have been studied: one is to measure the temperature change at a fixed location; the other one is to have the system carried by an unmanned drone to measure the temperature distribution of a selected area. The successful demonstrations show the potential applications of our wireless WGM sensor node in IoT. It is also worth noting that our sensing system is not limited to thermal sensing. With proper design, the packaged WGM sensor can have various functionalities, e.g., WGM magnetometer with high sensitivity and large bandwidth [19]. Another promising direction is to use a sensor array for monitoring multiple parameters simultaneously [27, 28].

## Competing financial interests

The authors declare no competing financial interests.

## Acknowledgements

This work is supported by the ARO grant No. W911NF1710189.

## Author contributions

L. Yang conceived the idea. X. Xu designed the hardware system and iOS software. W. Chen and X. Xu performed the sensing experiments with the help from Y. Li. G. Zhao fabricated the packaged WGM sensor. All authors contribute the analyzing and discussion of the experimental data. X. Xu, W. Chen, and L. Yang wrote the manuscript with contributions from all authors. L.Y. supervised the project.

# Figure Captions

**Fig. 1. Wireless WGM sensing system.** (a) Architecture of the wireless WGM sensing system. The light from a tunable single-mode distributed Bragg reflector (DBR) laser is used to probe a packaged whispering-gallery-mode sensor. The light coupled out of the sensor is sent to a photodetector with transmission amplifier (TIA). The ARM Cortex-M3 processor is responsible for controlling peripherals including laser current drive, thermo-electric cooler (TEC) controller, monitoring circuit and Wi-Fi unit. The sensing system is remotely controlled by an iOS app in a smartphone via the Wi-Fi unit. (b) Screenshot of the customized iOS app for wireless control of the sensing system. The system parameters, e.g., current, temperature, can be monitored and adjusted in real time. The transmission spectrum of the packaged sensor can also be acquired and analyzed in real time. (c) Photograph of the mainboard which integrates all the electronic components shown in (a). The size of the mainboard is about 124mm × 67mm.

**Fig. 2. Characterization of the wireless sensing system.** (a) Transmission spectrum of a packaged WGM sensor with frequency spanning of 450 GHz. Multiple resonances modes with different linewidths and polarizations can be obtained for sensing applications. (b) Transmission spectrum of a resonance mode (marked by a dashed box in (a)) with a Lorentzian fit. The quality factor is about $4.2 \times 10^5$. (c) Time trace of the linewidth of a resonance mode in the wireless sensing system.

**Fig. 3. Air temperature measurement.** The wireless sensing system was deployed outdoors to monitor the variation of air temperature from 8:30 AM to 8:30 PM on June 18[th], 2017 in St. Louis, MO, USA. Red circles denote the frequency shift of the selected resonance mode versus time, and the blue squares are the measurements of temperature change by commercial thermometer.

**Fig. 4. Aerial mapping of temperature distribution.** An unmanned drone was used to carry the wireless sensing system to measure the temperature distribution of a selected area in a city park of St. Louis. A commercial thermometer with Bluetooth connection was mounted together with the packaged sensor for comparison. (a) Photograph of an unmanned drone carrying the wireless sensing system (marked in the red dashed ellipse). (b) Frequency shift of the selected resonance mode when the drone flied in a selected loop where the starting and ending positions are marked. The resonance frequency at the starting position is set to be zero. The color bar represents the amount of frequency shift. The background image comes from Google Map. (c) Comparison of the measured frequency shift with the results from commercial thermometer. The increasing numbers denote the positions of measurements when the drone flied from the starting position to the ending position.

# FIGURES

**Fig. 1**

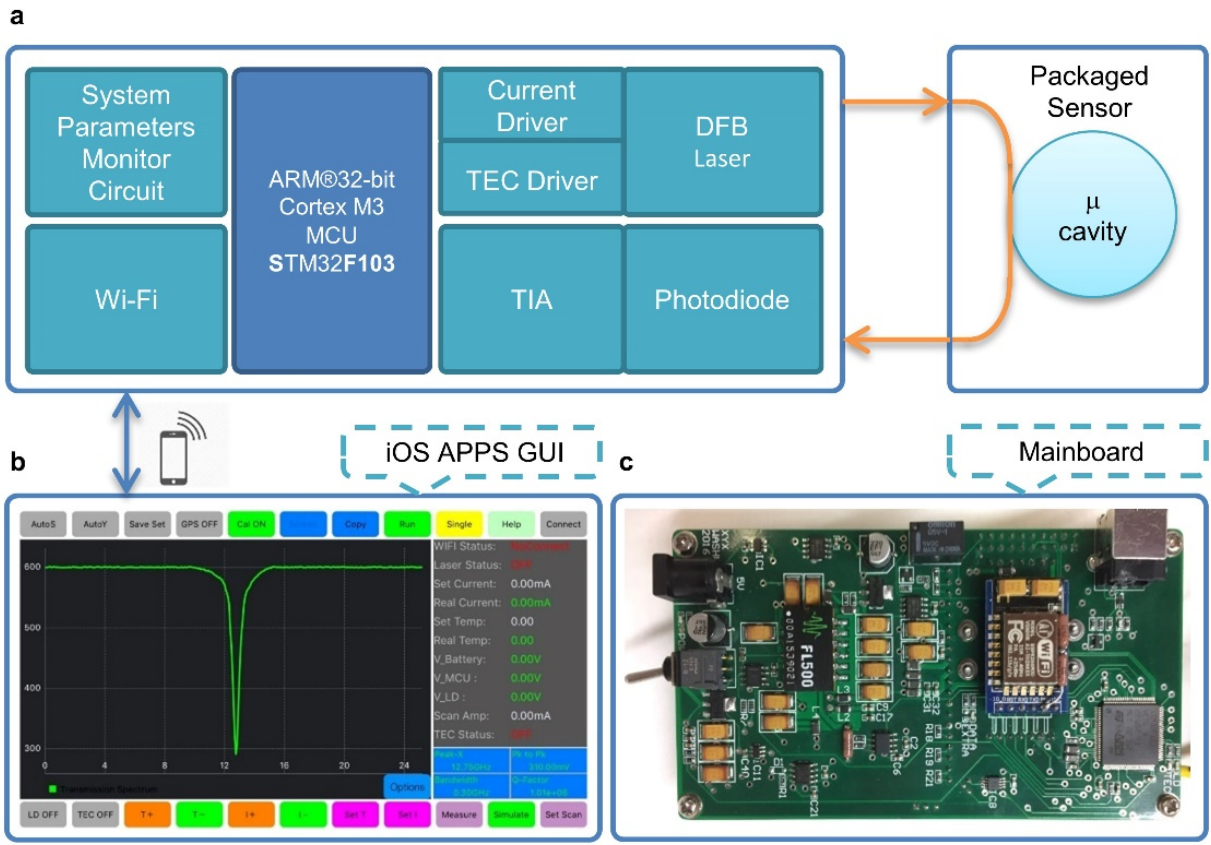

**Fig. 2**

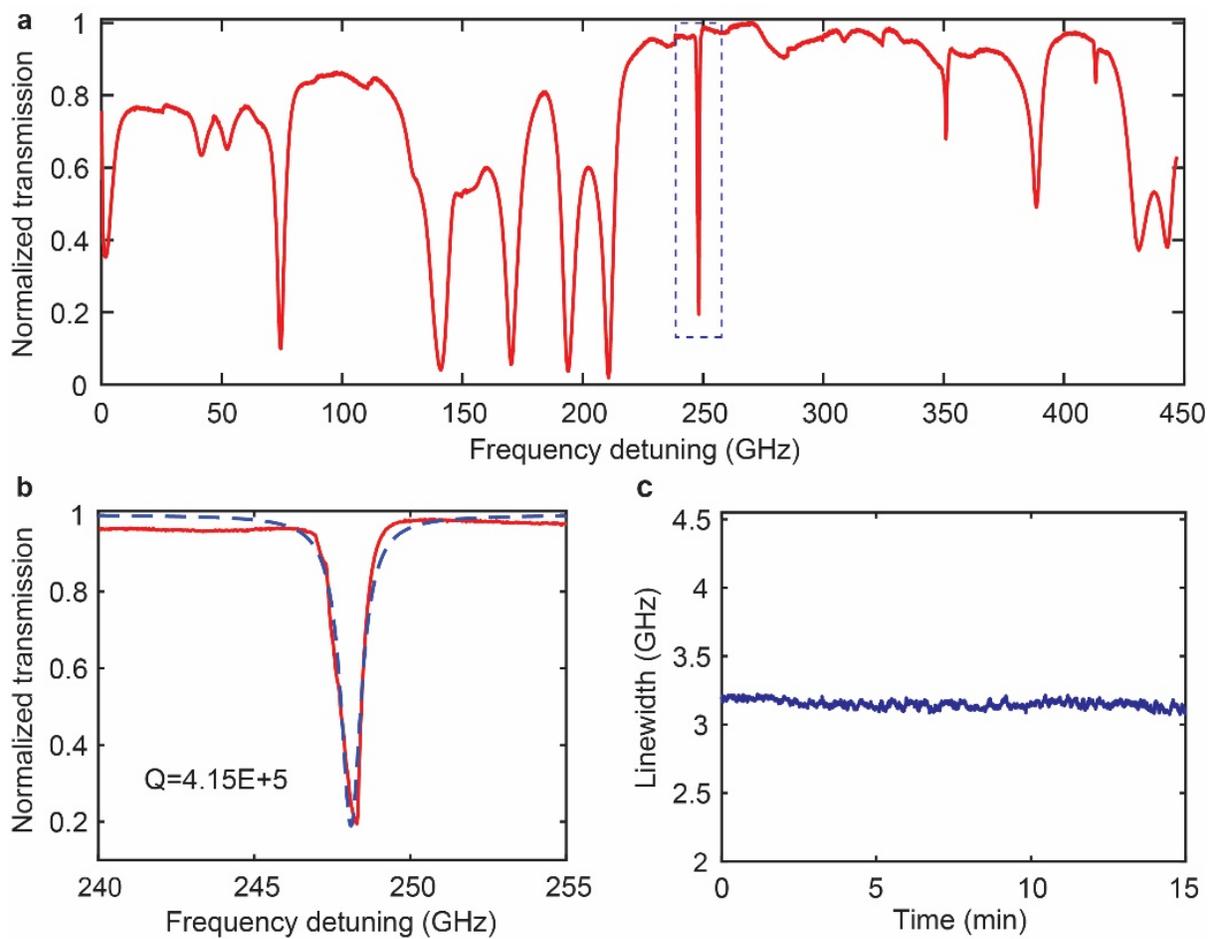

**Fig. 3**

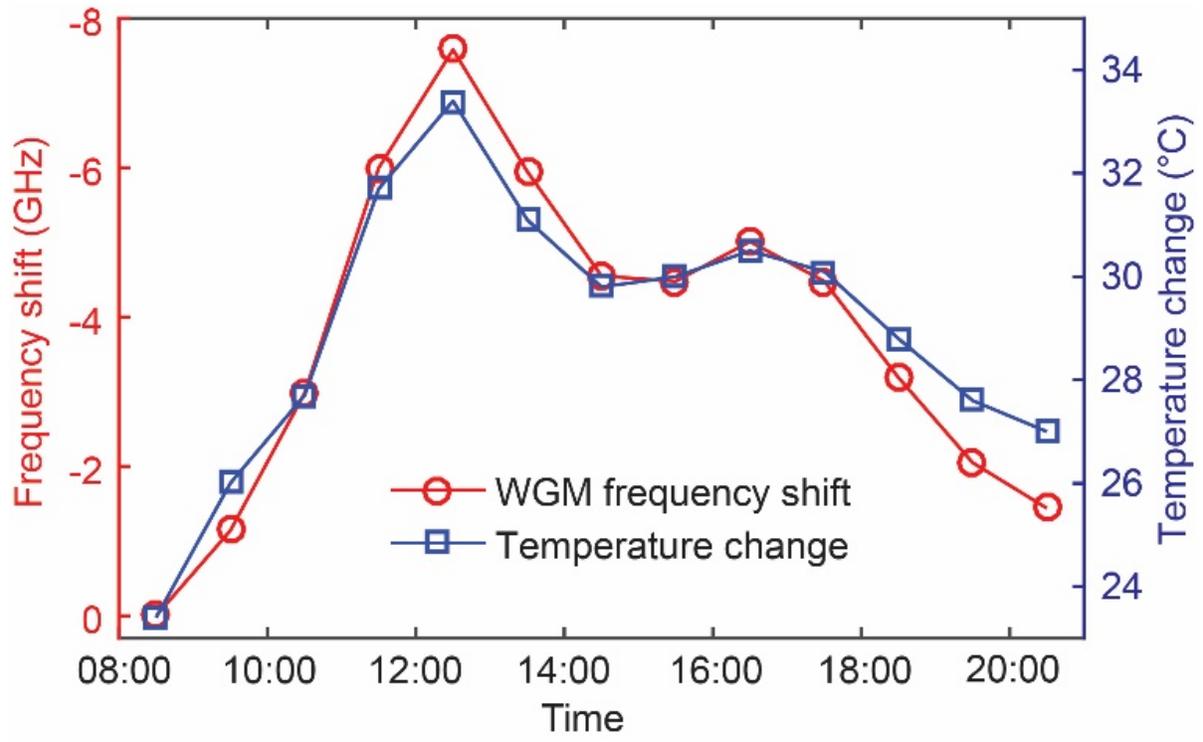

**Fig. 4**

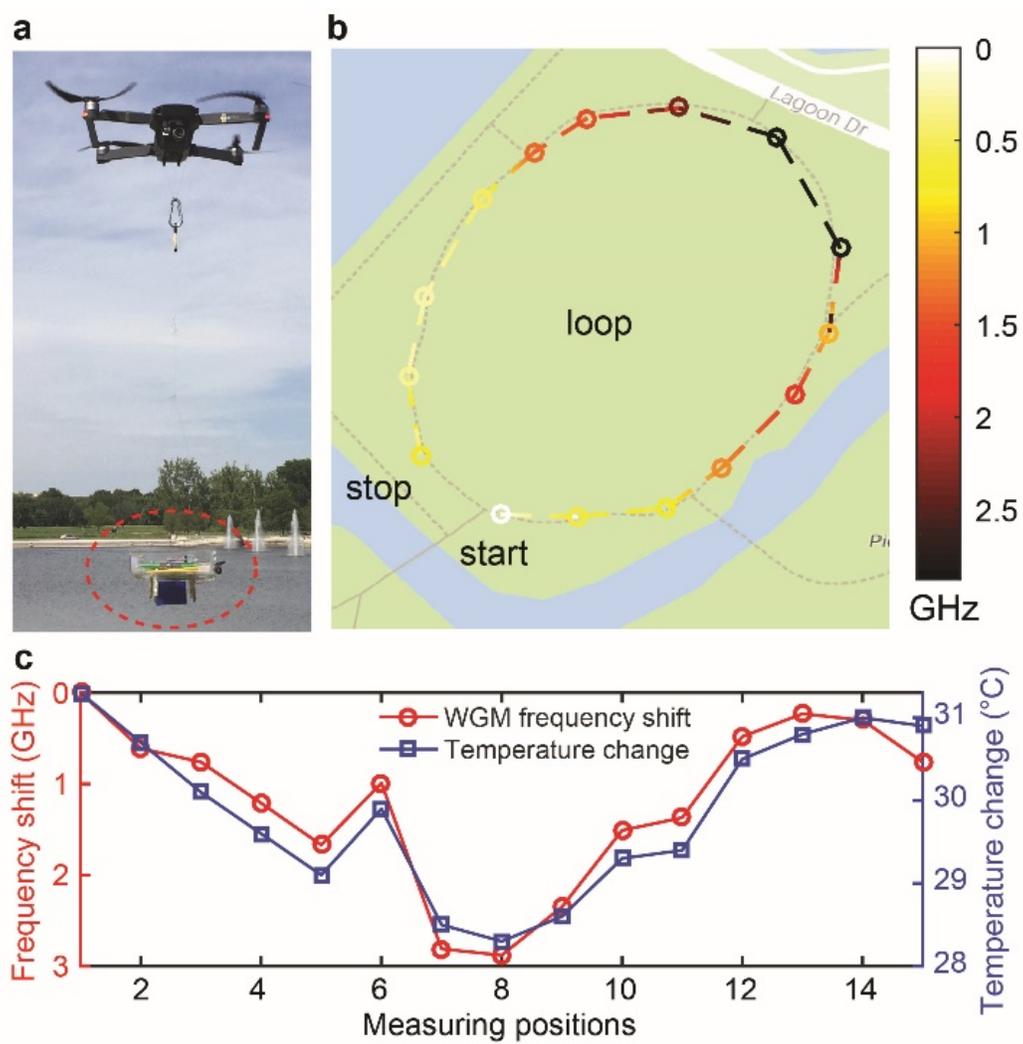